\documentclass[jcp, reprint, twocolumn, amsmath, amssymb, reprint, superscriptaddress]{revtex4-1}

\usepackage{graphicx,verbatim}
\usepackage{dcolumn}
\usepackage{bm}
\usepackage{sidecap}
\usepackage{braket}
\usepackage{color}
\usepackage{subcaption}

\begin{document}

\title{Communication:\ Hierarchical quantum master equation approach to vibronic reaction dynamics at metal surfaces}

\author{A. Erpenbeck}
\affiliation{
	Institute of Physics, Albert-Ludwig University Freiburg, \\
	Hermann-Herder-Strasse 3, 79104 Freiburg, Germany
}
\author{M.\ Thoss}
\affiliation{
	Institute of Physics, Albert-Ludwig University Freiburg, \\
	Hermann-Herder-Strasse 3, 79104 Freiburg, Germany
}

\date{\today}

\begin{abstract}
A novel quantum dynamical method to simulate vibronic reaction dynamics in molecules at metal surfaces is proposed. The method is based on the hierarchical quantum master equation approach and uses a discrete variable representation of the nuclear degrees of freedom in combination with complex absorbing potentials and an auxiliary source term. It provides numerically exact results for a range of models. By taking the coupling to the continuum of electronic states of the surface properly into account, nonadiabatic processes can be described and the effect of electronic friction is included in a nonperturbative and non-Markovian way. Illustrative application to models for desorption of a molecule at a surface and current-induced bond rupture in single-molecule junctions demonstrate the performance and versatility of the method.
\end{abstract}

\maketitle

\paragraph*{\textbf{Introduction:}}

\begin{figure}[b!]
	\hspace*{1cm}\raggedright a) \hspace*{3.cm} b)\\
	\centering
	\begin{subfigure}[l]{0.133\textwidth}
		\vspace*{-0.33cm}
		\includegraphics[width=\textwidth]{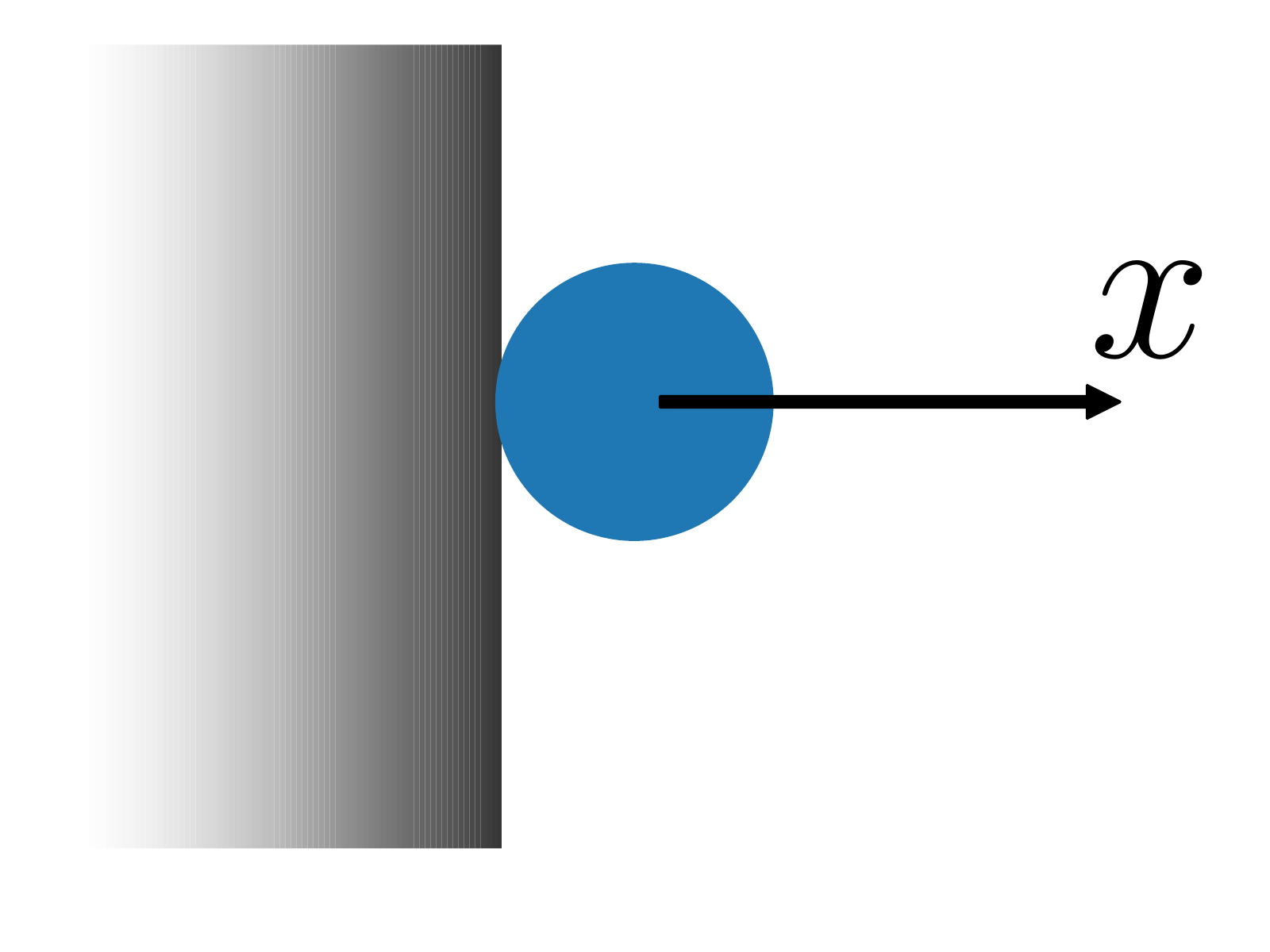}
	\end{subfigure}
	\hspace*{1cm}
	\begin{subfigure}[l]{0.133\textwidth}
		\vspace*{-0.33cm}
		\includegraphics[width=\textwidth]{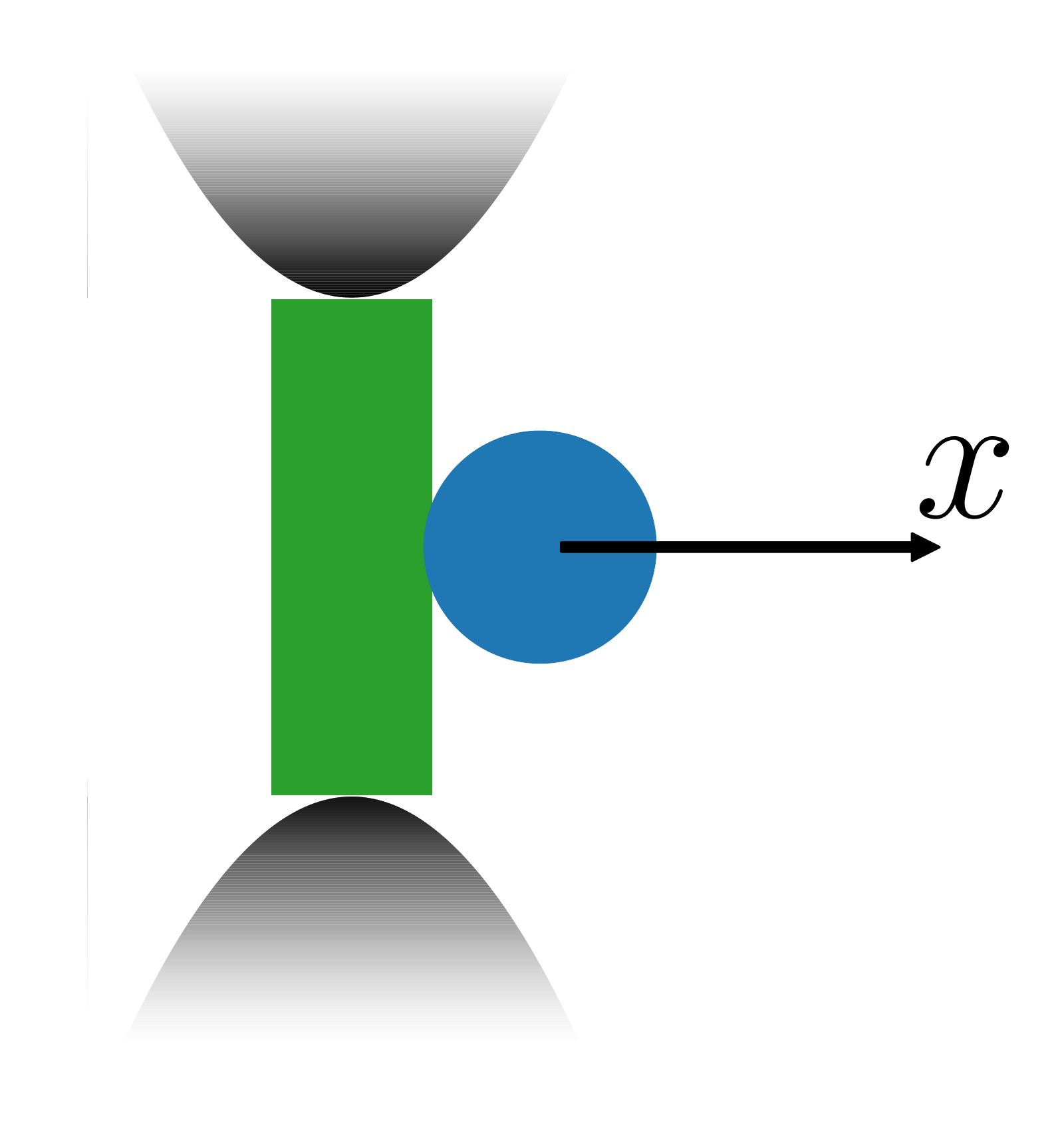}
	\end{subfigure}
	\captionsetup{font=small,labelfont=bf, justification=centerlast, format=plain}
 	\caption{ \bf \scriptsize 
		Molecule-surface systems considered.
		a:  Molecule adsorbed on a surface.
		b: Single-molecule junction. 
		{\color{black}The gray triangular areas represent the two leads. 
		The molecule consists of a backbone bridging the leads (green rectangle) and a side-group (blue circle), which may detach as a consequence of a current flowing across the molecule.}
	}
	\label{fig:setups}
\end{figure}

Understanding the dynamics of molecules interacting with metal surfaces is of great importance in physics, chemistry, and technology. Examples of dynamical processes include  
the chemisorption and desorption of molecules at surfaces (see Fig.\ \ref{fig:setups}a), the scattering of molecules from surfaces, reactive and catalytic processes as well as electron transport through molecules in STM setups or single-molecule junctions (see Fig.\ \ref{fig:setups}b).
An important aspect in these systems is the influence of the electrons of the surface on the dynamics of the molecule. The coupling to the continuum of electronic states of the surface results in electronic and vibrational relaxation processes including electron-hole pair creation and can also cause strong nonadiabatic effects because of a breakdown of the Born-Oppenheimer approximation.

The theoretical description of such scenarios represents a challenging task, in particular for reactive processes.\cite{Guo1999, Tully2000, Saalfrank2006, Jiang2019}
A variety of approaches have been developed and applied in this context. Particularly popular are mixed quantum classical methods, which treat the nuclear degrees of freedom (DOFs) in the classical approximation, including 
Ehrenfest-type approaches, the surface hopping scheme or Langevin equations.\cite{Tully1971, Cheng2007, Shenvi2009, Fischer2011, Dzhioev2011, Dzhioev2013, Galperin2015, Miao2017, Erpenbeck2018, Lue2019} Furthermore, a variety of 
density matrix based schemes have been applied, which use either classical or pertubative approximations.\cite{Saalfrank1996, Gao1998, Guo2000, Leathers2009} 
The concept of electronic friction has been employed in several of these approaches.\cite{Head‐Gordon1995, Wodtke2004, Miao2017, Maurer2017, Dou2018, Rittmeyer2018}

Despite the impressive success of these methods, the inherent approximations limit their applicability to certain parameter regimes. In this paper, we propose a novel approach based on the hierarchical quantum master equation method (HQME, also referred to as hierarchical equations of motion (HEOM)),\cite{Tanimura1989, Tanimura2006, Jin2008, Zheng2012, Haertle2013a,Schinabeck2016} which describes the electronic and the nuclear DOFs nonperturbatively on 
{\color{black} a quantum level.}
For a range of models of molecule-surface systems, it is {\color{black}capable of providing} numerically exact results for the coupled electronic-vibrational (vibronic) dynamics.
The novel approach combines various well-established techniques, namely the HQME method with a discrete variable representation (DVR) for the nuclei in combination with a complex absorbing potential (CAP) and {\color{black} an associated} source term.
Another approach, which also allows, in principle, a numerically exact treatment of vibronic reaction dynamics at metal surfaces is the multilayer multiconfiguration time-dependent Hartree (ML-MCTDH) method in second quantization representation.\cite{Wang2009, Manthe2017}\\

\paragraph*{\textbf{Model and method:}} \label{sec:HQME}

	The method employs a system-bath framework to describe the quantum dynamics of a molecule coupled to a metal surface. Thereby, the system comprises the electronic and selected nuclear DOFs (reaction modes) of the molecule. The baths, which represent the metal surfaces, are modeled as electron reservoirs. Correspondingly, the Hamiltonian adopts the form
	$H = H_{\text{S}} + H_{\text{B}} + H_{\text{SB}}$, 
	with $H_{\text{S}}$ being the Hamiltonian of the system, $H_{\text{B}}$ the Hamiltonian describing the electronic baths, and the coupling given by $H_{\text{SB}}$. 
	{\color{black}The molecular Hamiltonian assumes the general form
	\begin{eqnarray}
		H_{\text{S}}	&=& 	T_{\text{nuc}} + \sum_{N n n'} \ket{N,n} \mathcal{W}_{Nnn'}(\mathbf{x}) \bra{N,n'},
	\end{eqnarray}
	where $T_{\text{nuc}}$ represents the kinetic energy of the nuclei with associated coordinates $\textbf{x}$.
	$\ket{N,n}$ are the relevant diabatic electronic states of the molecule, where $N$ denotes the overall charge of the molecule ($N=0, \pm1, \dots$) and $n$ is the quantum number corresponding to the state.
	Thus, e.g., $\ket{N=-1,n=0}$ denotes the electronic ground-state of the molecular anion, while $\ket{N=0,n=1}$ is the first excited state of the neutral molecule.
	The diagonal matrix elements $\mathcal{W}_{Nnn}(\textbf{x})$ are the diabatic potential energy surfaces (PESs) of state $n$, whereas the non-diagonal elements $\mathcal{W}_{Nnn'}(\textbf{x})$ with $n\neq n'$ describe the coupling between the different states $n$ and $n'$. }
	The electrons in the metal surface are modeled as effectively noninteracting baths, where $c_k^{\dagger}/c_k$ denote the electronic creation/annihilation operator associated to state $k$ of bath $B_i$ with energy $\epsilon_k$.
	\begin{eqnarray}
		H_{\text{B}}	&=&	\sum_{B_i}\sum_{k\in B_i} \epsilon_k c_k^\dagger c_k, \label{eq:H_B}
	\end{eqnarray}
	{\color{black}The coupling between the molecule and the surfaces is described by the Hamiltonian
	\begin{eqnarray}		
		H_{\text{SB}}	&=&	\sum_{B_i k}\sum_{Nnm} \left( V_{knm}^N(\textbf{x}) c_k^\dagger S_{Nnm}  + \text{h.c.} \right) , \label{eq:H_coupl}
	\end{eqnarray}
	where $S_{Nnm}=\ket{N,n}\bra{N+1,m}$ is a generalized annihilation operator which encodes the transition between two molecular states which differ in the electron number by one.
	This} coupling between the molecule and the surfaces in Eq.\ (\ref{eq:H_coupl}) gives rise to the spectral density of bath $B_i$,
	{\color{black}
	\begin{eqnarray}
		\Gamma_{B_i nm}^N(\epsilon, \textbf{x}) = 2\pi\sum_{k\in B_i \atop n'} V^N_{knn'}(\textbf{x}) V^{N*}_{k'n'm}(\textbf{x}) \delta(\epsilon-\epsilon_k) ,\  \label{eq:spectral_func} 
	\end{eqnarray}
	} 
	\hspace*{-0.32cm}
	which may depend on the nuclear DOFs.
	Notice, that even though we apply a time-independent formulation in this paper, a generalization to account for the effect of time-dependent energies and coupling strengths is in principle straightforward. Moreover, an extension to also incorporate bosonic reservoirs describing the phonons of the metal surfaces or less important vibrational modes of the molecule is also possible but not considered in the present paper.

	The coupled vibronic dynamics is described using the HQME method.
	Generally, the HQME method is a reduced density matrix approach originally developed by Tanimura and Kubo in the context of molecular relaxation dynamics,\cite{Tanimura1989, Tanimura2006} which describes the dynamics of a quantum system coupled to a bath.
	It extends perturbative master equation methods by including higher-order contributions and non-Markovian effects and can provide numerically exact results.
	For a detailed account of the method in the context of modeling electronics baths we refer to Refs.\ \onlinecite{Jin2008, Zheng2012, Haertle2013a,Schinabeck2016}. Here, we extend the HQME method to study vibronic dynamics at metal surfaces beyond the harmonic approximation for the nuclei.

	Within the HQME framework, the influence of the electronic baths as modeled by Eqs.\ (\ref{eq:H_B}) and (\ref{eq:H_coupl}) is characterized by the bath correlation function
	\begin{eqnarray}
		C_{B_i{\color{black}nm}}^{{\color{black}N}\pm}(t,t', \textbf{x})	
		&=&	
								    \int d\epsilon \ e^{\pm \frac{i}{\hbar}\epsilon (t-t')} \Gamma_{B_i{\color{black}nm}}^{{\color{black}N}}(\epsilon, \textbf{x}) f(\pm\epsilon, \pm\mu_{B_i}) \ , \nonumber \\
								    \label{eq:correlation_func_general}
	\end{eqnarray}
	with the Fermi function $f(\epsilon, \mu) = \left( 1 + \exp(\beta(\epsilon-\mu))  \right)^{-1}$. Here, $\beta=\frac{1}{k_{B}T}$ where $k_B$ is the Boltzmann constant, $T$ the temperature, and $\mu$ the chemical potential.
	In order to obtain a closed set of equations within the HQME approach, it is expedient to represent the bath correlation function as a sum over exponentials,\cite{Jin2008}
	\begin{eqnarray}
		C_{B_i{\color{black}nm}}^{{\color{black}N}\pm}(t,t', \textbf{x})
			&\equiv& 	
					\sum_{q=1}^\infty     \eta_{B_i{\color{black}nm}q\pm}^{{\color{black}N}}(\textbf{x})  e^{-\gamma_{B_i{\color{black}nm}q\pm}^{{\color{black}N}} (t-t')} \ .  \label{eq:decomposition_correlation_function} 
	\end{eqnarray}
	Common approaches for obtaining the parameters $\eta_{B_i{\color{black}nm}q\pm}^{{\color{black}N}}$ and $\gamma_{B_i{\color{black}nm}q\pm}^{{\color{black}N}}$ include the Matsubara\cite{Mahan, Tanimura2006, Jin2008} and the Pade decomposition,\cite{Hu2010, Hu2011} yet more sophisticated schemes exist.\cite{Popescu2015, Tang2015, Popescu2016, Ye2017, Erpenbeck_RSHQME}
	Notice, that this scheme also works with modified definitions of the bath correlation function, which can facilitate the description of effects such as time- and position-dependent molecule-lead coupling strengths.\cite{Erpenbeck2018}

	The HQME method employs a set of auxiliary density operators $\rho_{j_1 \dots j_n}^{(n)}$, which incorporate the electronic and nuclear DOFs of the system and which obey the equation of motion (EOM)	\begin{eqnarray}
	 	\frac{\partial}{\partial t} \rho_{j_1 \dots j_n}^{(n)}(t) 	&=& 
		\left[-\frac{i}{\hbar} \left( \mathcal{L}_\text{S} + \mathcal{F}\right) -\left( \sum_{m=1}^n \gamma_{j_m} \right)  \right] \rho_{j_1 \dots j_n}^{(n)}(t)
		\nonumber \\&&
		-i \sum_{m=1}^n (-1)^{n-m} \mathcal{C}_{j_m} \rho_{j_1\dots j_{m-1} j_{m+1} \dots j_n}^{(n-1)}(t) \nonumber \\&&
		-\frac{i}{\hbar^2} \sum_{j} A^{\overline{\sigma_{j}}}_{\nu_j} \rho_{j_1 \dots j_n j}^{(n+1)}(t) \ .
		\label{eq:EQM_nth_tier}
	\end{eqnarray}
	This notation uses the multi-index $j_i = (B_i, {\color{black}N, n, m}, q_i, \sigma_i)$, where $B_i$ labels the different baths, {\color{black}$n$, $m$ are many-particle states of the molecule associated to the electron number $N$}, $\sigma_i = \pm 1$ and $q_i$ is the pole-index related to the decomposition in Eq.\ (\ref{eq:decomposition_correlation_function}). Moreover, $\overline{\sigma} = -\sigma$ and $\mathcal{L}_\text{S} O= [H_\text{S}, O]$.
	$\rho^{(0)}$ is the reduced density operator of the system, the higher-tier auxiliary density operators $\rho_{j_1 \dots j_n}^{(n)}$ encode the influence of the electronic reservoirs on the system dynamics.	
	The operators $A^{{\sigma}}_{\nu}$ and $\mathcal{C}_{j}$ couple the $n^{\text{th}}$-tier to the $(n+1)^{\text{th}}$- and $(n-1)^{\text{th}}$-tier auxiliary density operators, 
	{\color{black}
	\begin{subequations}
		\begin{eqnarray}
			A^{N\sigma}_{nm} \rho^{(n)}(t)	&=&	\left\lbrace V_{knm}^{N\sigma}(\textbf{x}) S_{Nnm}^\sigma\ ,\ \rho^{(n)}(t)\right\rbrace_{(-)^n} , \\
%
			\mathcal{C}_{j} \rho^{(n)}(t)	&=&	\eta_{j}^\sigma(\textbf{x})\ S_{Nnm}^\sigma \rho^{(n)}(t)
							\\&& - (-1)^n \rho^{(n)}(t) S_{Nnm}^\sigma \eta_{j}^{\overline{\sigma}*}(\textbf{x}) , \nonumber 
		\end{eqnarray}
	\end{subequations}
	}
	leading to a hierarchy of EOMs.
	Thereby, we employ the notation {\color{black}$S_{Nnm}^- \equiv S_{Nnm}$ and $S_{Nnm}^+ \equiv S_{Nnm}^\dagger$} as well as $\eta_{j}^- \equiv \eta_{j}$ and $\eta_{j}^+ \equiv \eta_{j}^*$. 
	The HQME approach is formally exact given that $H_{\text{B}}$ and $H_{\text{SB}}$ assume the form given in Eqs.\ (\ref{eq:H_B}) and (\ref{eq:H_coupl}). For applications, the hierarchy needs to be truncated in a suitable manner. Also, only a finite number of poles characterizing the bath can be taken into account. For details of the convergence properties of the HQME method we refer to Refs.\ \onlinecite{Tanimura1991, Xu2005, Shi2009, Hu2010, Schinabeck2018, Dunn2019}.
	Notice, that all operators entering the EOMs (\ref{eq:EQM_nth_tier}) as well as the (auxiliary) density operators may also act on the nuclear DOFs. 
	
	In order to facilitate a description of the nuclear dynamics based on generic PESs, we employ a DVR.\cite{Tannor,Colbert1992} Furthermore, to avoid finite size effects, such as reflections at the boundary of the grid, we use a CAP, which absorbs the parts of the wavefunction beyond a certain distance from the surface.

	\begin{figure}[tb!]
		\raggedright a)\\
		\hspace*{-0.4cm}
		\includegraphics[width=0.45\textwidth]{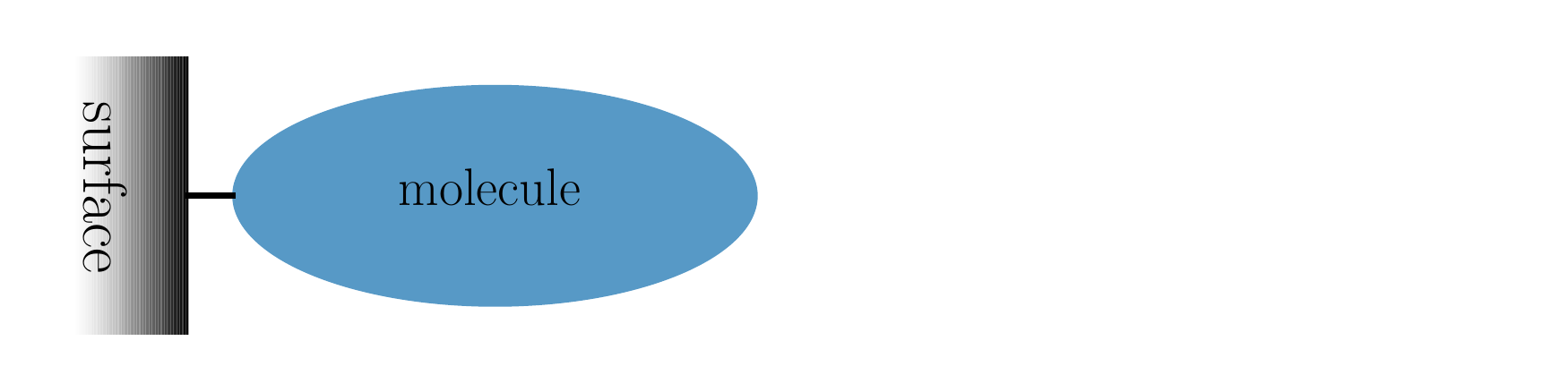}\\
		\raggedright b)\\
		\hspace*{-0.4cm}
		\includegraphics[width=0.45\textwidth]{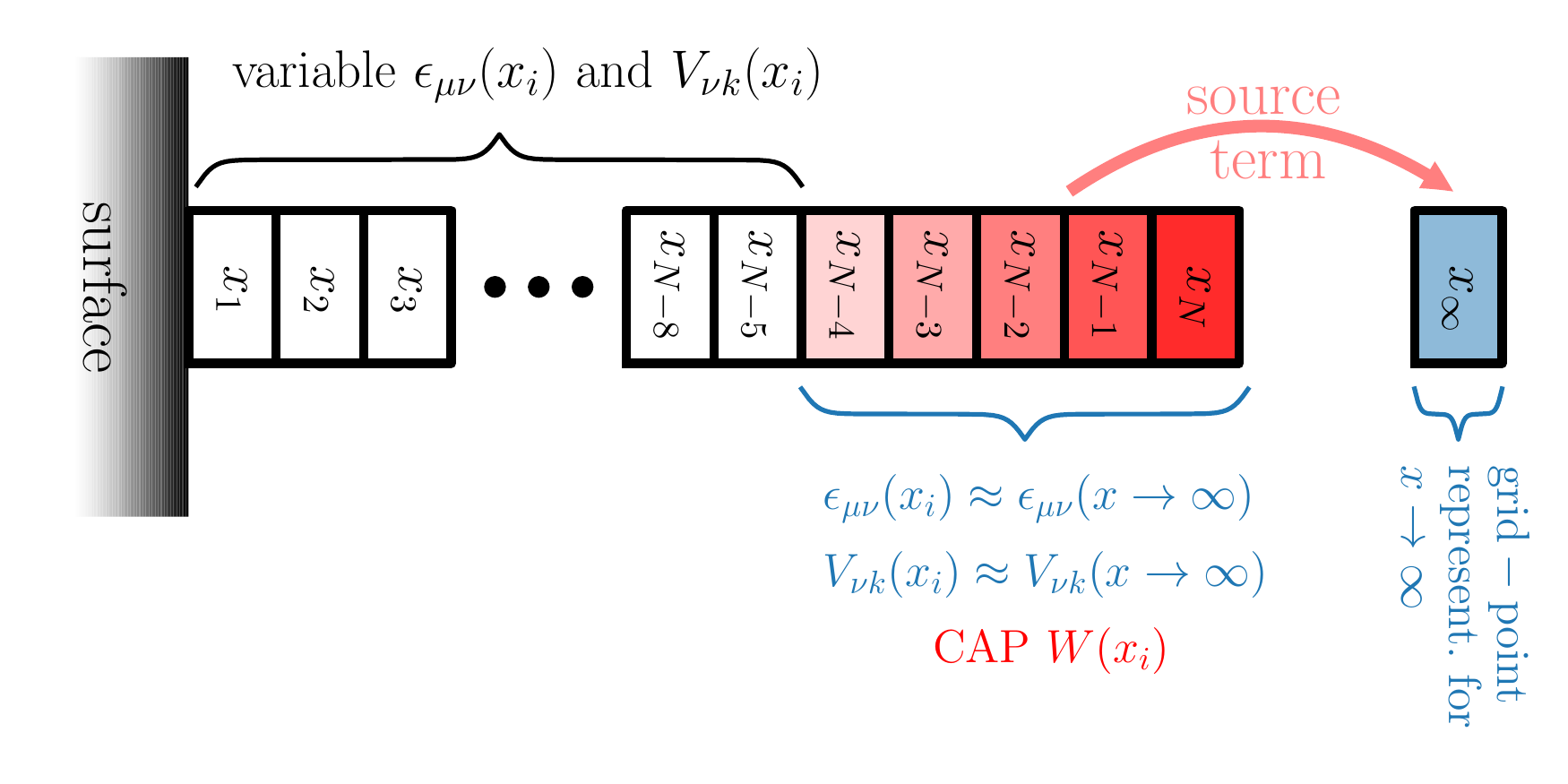}
		\captionsetup{font=small,labelfont=bf, justification=centerlast, format=plain}
		\caption{ \bf \scriptsize 
			Visualization of the methodological concepts. 
			a: Real-space representation of the molecule surface model.
			b: Sketch of the DVR and the source term employed. The individual rectangles represent the different DVR grid points. The red shading emphasizes the action of the CAP. The source term maps the probability absorbed by the CAP to the auxiliary grid point $x_\infty$, which is representative of large distances from the surface.
		}
		\label{fig:concept}
	\end{figure}
	As the application of a CAP within a many-body approach leads to problems associated with the conservation of the particle number,\cite{Selsto2010, Kvaal2011, Prucker2018} we additionally introduce {\color{black} an associated} source term, which maps the probability absorbed by the CAP to additional grid points $\mathbf{x}_\infty$, which are representative of large distances from the surface. This is motivated by the assumption that the potentials 
	{\color{black}$\mathcal{W}_{Nnn'}(\textbf{x})$ and $V_{knm}^N(\textbf{x})$ }
	are constant at large distances from the surface.
	As such, the electronic properties are described correctly even for large displacements from the surface. 
	{\color{black} This is essential in situations such as chemical decomposition at surfaces, where a part of the adsorbate remains attached to the surface, or nondestructive current-induced dissociation in molecular junctions, where a side-group detaches from the molecule which still bridges the leads (see Fig.\ \ref{fig:setups}b).}
	{\color{black} A meaningful description of these scenarios, e.g. the electronic properties of the part of the adsorbate remaining on the surface or the current flowing through the molecular junction, must keep track of the properties associated to the part of the probability lost via the CAP. The additional source term maps the probability absorbed by the CAP to representative grid points.}
	This strategy, which was used in a similar fashion in Ref.\ \onlinecite{Prucker2018}, is encoded in the operator $\mathcal{F}$ in Eq.\ (\ref{eq:EQM_nth_tier}), defined as
	\begin{eqnarray} 
		\mathcal{F}(\rho_{j_1 \dots j_n}^{(n)}(t))	&=&	   - i \lbrace W(\textbf{x}), \rho_{j_1 \dots j_n}^{(n)}(t) \rbrace \label{eq:7:Lindblad_explicit} \label{eq:Lindblad_Op}\\&& 
								\hspace*{-1cm}
									  + 2 i \left( \sum_{\textbf{x}_i} W(\textbf{x}_i) \braket{\textbf{x}_i| \rho_{j_1 \dots j_n}^{(n)}(t) | \textbf{x}_i} \right) \ket{\textbf{x}_\infty}\bra{\textbf{x}_\infty} , \nonumber
	\end{eqnarray}
	{\color{black}where $\textbf{x}_i$ denote the DVR grid points.}
	It incorporates the CAP $W(\textbf{x})$ (see first summand in Eq.\ (\ref{eq:Lindblad_Op})) and the associated source term (second summand in Eq.\ (\ref{eq:Lindblad_Op})).
%
%
	{\color{black} $\mathcal{F}$ represents an operator which can be expressed in Lindblad form. It is important to emphasizes, though, that it contrast to master equation approaches, which employ the Lindblad framework, in the present context, the operator $\mathcal{F}$ only implements the CAP, while the physical dynamics is described by the HQME.}
	We note in passing, that the emergence of the operator $\mathcal{F}$ in the EOMs (\ref{eq:EQM_nth_tier}) can be derived revisiting the derivation of the HQME while employing a formal solution for the system dynamics including the CAP and the associated source term. The concept represented by the operator $\mathcal{F}$ is visualized for a single surface in Fig.\ \ref{fig:concept}.\\

\paragraph*{\textbf{Results:}} \label{sec:results}
	\begin{figure}[tb!]
		\centering
		\includegraphics[width=0.9\linewidth]{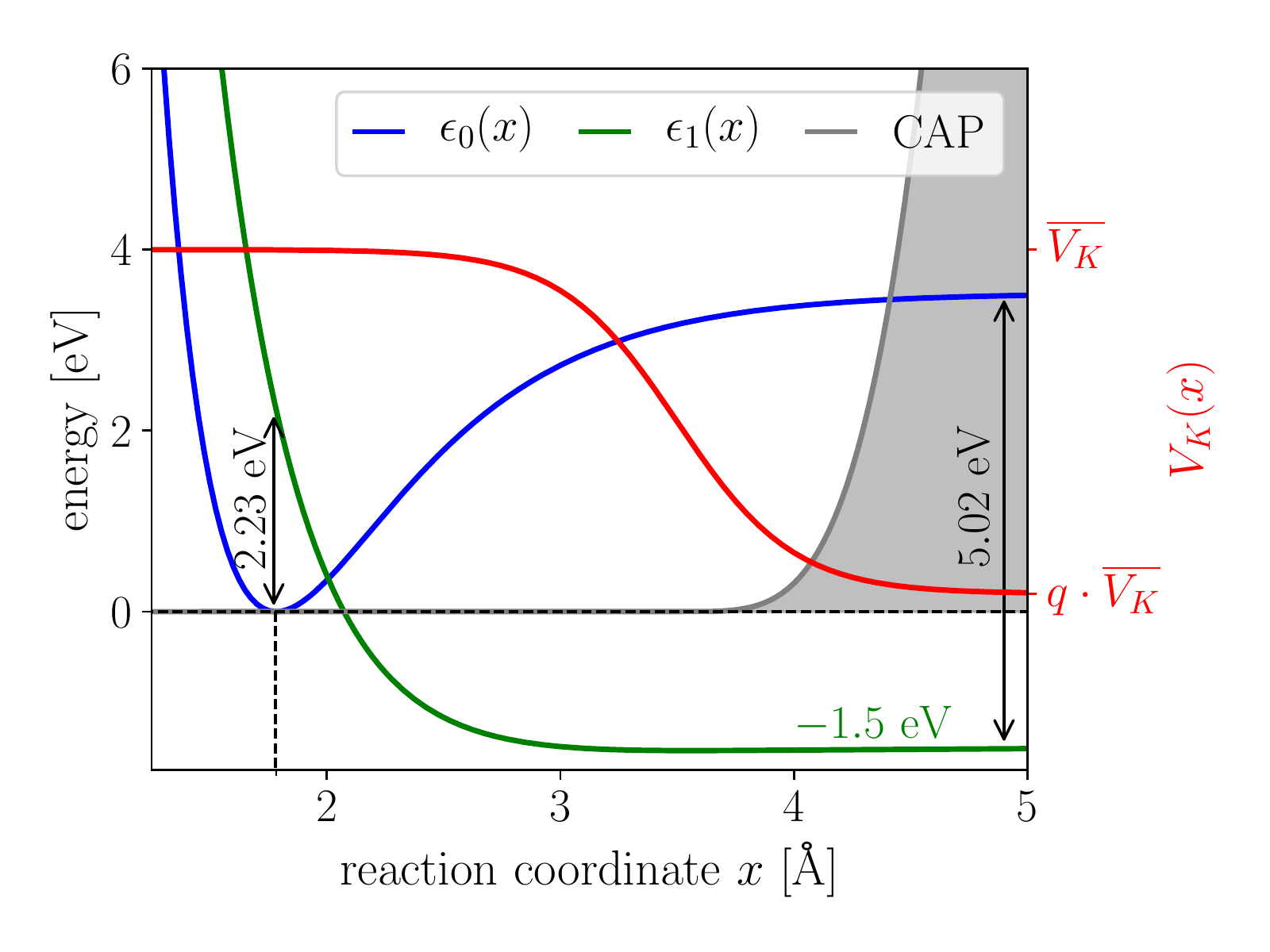}
		\captionsetup{font=small,labelfont=bf, justification=centerlast, format=plain}
		\caption{\bf \scriptsize 
		      Potentials $\epsilon_0(x)$, $\epsilon_1(x)$, as introduced in Eqs.\ (\ref{eq:V_0}) and (\ref{eq:V_d}), molecule-surface coupling strength $V_{k}(x)$ as defined in Eq.\ (\ref{eq:def_mol_lead_coupling_strength}) and CAP $W(x)$.}
		\label{fig:potentials0}
	\end{figure}

	\begin{figure*}[tb!]
		\begin{subfigure}[l]{0.325\textwidth}
			\raggedright a)\\
			\includegraphics[width=\textwidth]{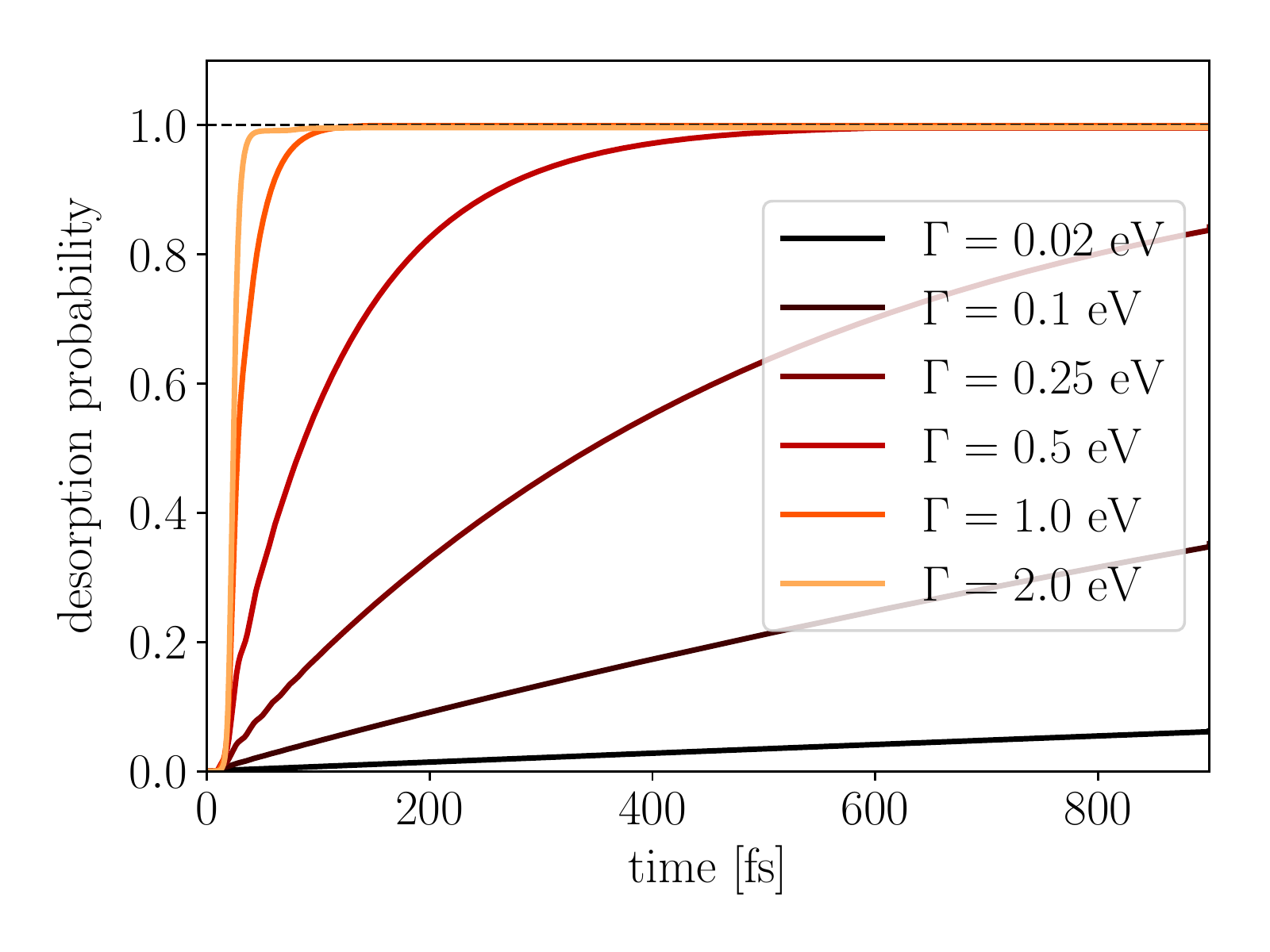}
		\end{subfigure}
		\begin{subfigure}[l]{0.325\textwidth}
			\raggedright b)\\
			\includegraphics[width=\textwidth]{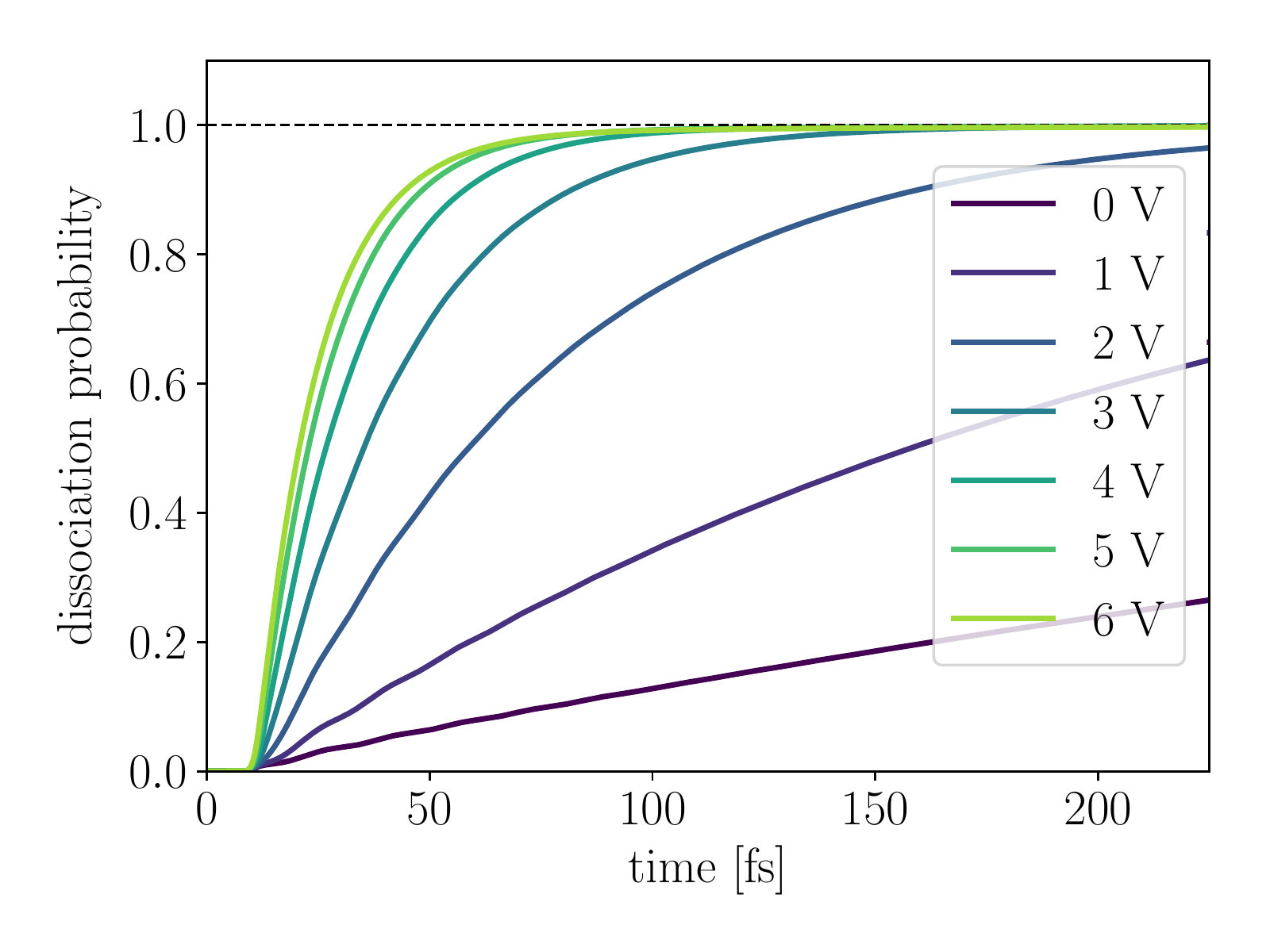}
		\end{subfigure}
		\begin{subfigure}[l]{0.325\textwidth}
			\raggedright c)\\
			\includegraphics[width=\textwidth]{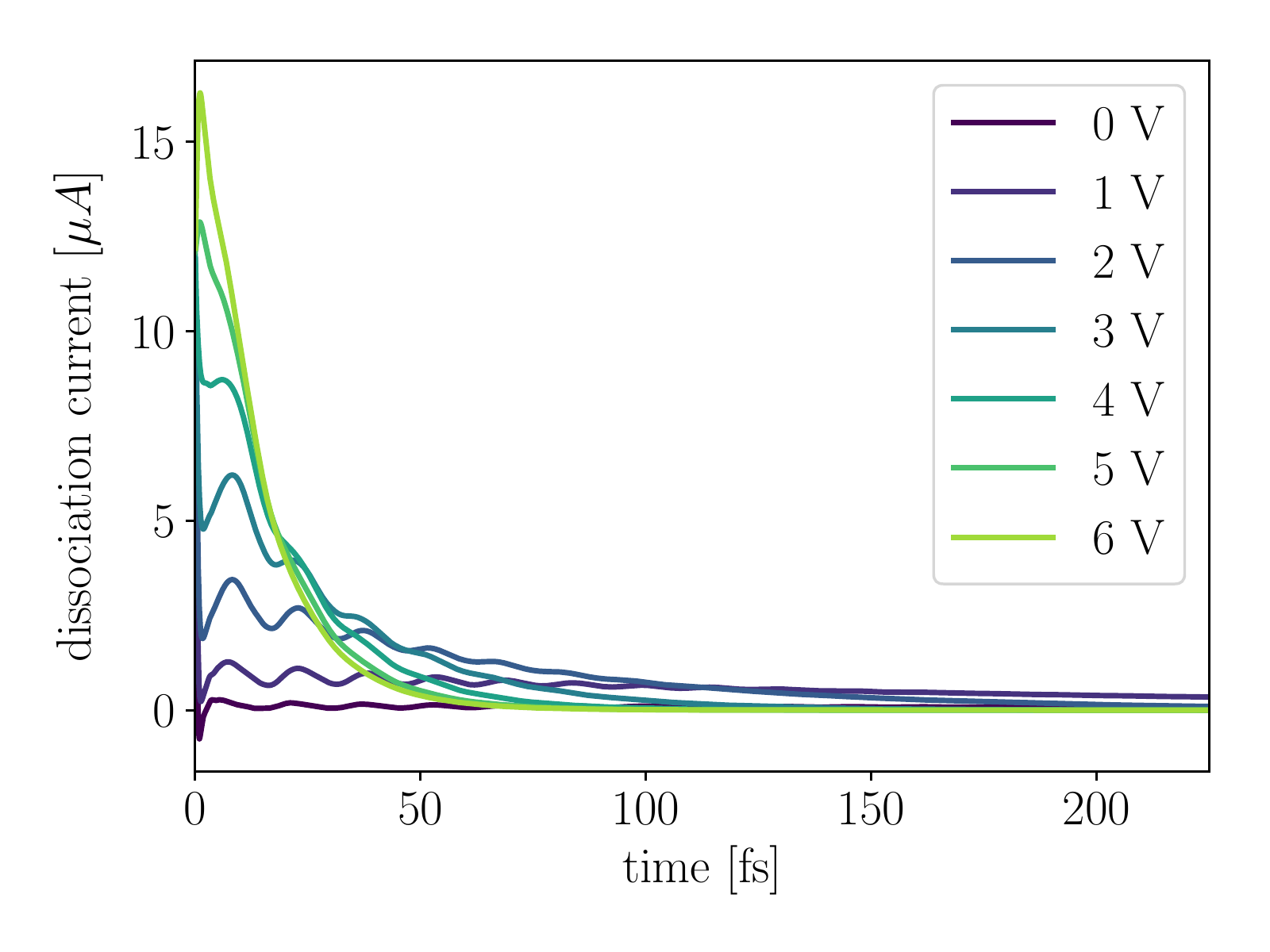}
		\end{subfigure}
 		\captionsetup{font=small,labelfont=bf, justification=centerlast, format=plain}
		\caption{ \bf \scriptsize
		a: Probability for desorption from a surface as a function of time for different coupling strengths $\Gamma$.
		b: Probability for the detachment of a side-group from a molecular junction as a function of time for different applied bias voltages for $\Gamma=0.1$ eV.
		c: Current through a molecular junction as a function of time while allowing for current-induced dissociation for different applied bias voltages for $\Gamma=0.1$ eV.
		}
		\label{fig:potentials}
	\end{figure*}
	As an example to illustrate the methodology we consider a basic model for vibronic dynamics of a molecule at a surface. 
	{\color{black} To this end, we consider an idealized molecule using the Anderson-Holstein model which}
	includes a single electronic 
	{\color{black} orbital}
	and a single reaction mode $x$ described by the Hamiltonian
	\begin{eqnarray}
		H_{\text{S}}	&=&	\frac{p^2}{2m} + {\color{black} \epsilon_0(x) + \Big(\epsilon_1(x)-\epsilon_0(x) \Big) d^\dagger d } . \label{eq:Hamiltonian_results}
	\end{eqnarray}
	{\color{black} Here, $d^\dagger$ and $d$ are the electronic creation and annihilation operators of the electronic orbital, respectively.}
	The molecule described by this model can be in a neutral state (electronic {\color{black} orbital} is unpopulated) or in an anionic state (electronic {\color{black} orbital} is populated).
	The PES of the neutral molecule is assumed to be a binding Morse-potential of the form
	\begin{eqnarray}
	      \epsilon_0(x) 	&=&	D_e \cdot\left( e^{-a(x-x_0)} -1 \right)^2 + c , \label{eq:V_0}
	\end{eqnarray}
	while the PES of the charged molecule adopts a non-binding form,
	\begin{eqnarray}
		\epsilon_1(x) &=&	D_1 \cdot e^{-2\cdot a' (x-x_0')} - D_2 \cdot e^{-a' (x-x_0')}  + V_\infty . \label{eq:V_d}
	\end{eqnarray}
	In the numerical results presented below, we have used the parameters $m=1$ amu (atomic mass units), $D_e = 3.52$ $\text{eV}$,  $x_0=1.78$ \AA, $a=1.7361$~\AA$^{-1}$, $D_1=4.52$ $\text{eV}$, $D_2=0.79$ $\text{eV}$, $x_0'=1.78$ \AA\ and $a'=1.379$ \AA$^{-1}$. The parameter $c= -147$~meV is chosen such that the nuclear ground-state has the energy $0$ eV. $V_\infty$ is set to $-1.5$ eV. 
	Moreover, we describe the molecule-surface coupling 
	{\color{black} by the Hamiltonian
	\begin{eqnarray}
	      H_{\text{SB}}	&=&	\sum_{k} \left( V_{k}(\textbf{x}) c_k^\dagger d  + \text{h.c.} \right)  ,
	\end{eqnarray}
	where the generalized annihilation operator from Eq.\ (\ref{eq:H_coupl}) is replaced by $d$ and with}
	\begin{eqnarray}
		V_{k}(x)    &=&    \overline V_{k} \cdot \left( \frac{1-q}{2} \left[ 1-\tanh\left(\frac{x-\tilde x}{\tilde a} \right) \right] + q \right), \ \label{eq:def_mol_lead_coupling_strength}
	\end{eqnarray}
	{\color{black} whereby $\overline V_{k}$ is the maximal coupling strength}. The other parameters are $q=0.05$, $\tilde a=0.5$ \AA\ and $\tilde x = 3.5$ \AA. 
	The {\color{black}function form} chosen for the dependence of the coupling on the nuclear coordinate $x$ describes a decrease of the coupling strength for larger values of $x$, as is to be expected, e.g., for desorption of a molecule from a surface. 
	The corresponding potentials and the system-bath coupling are visualized in Fig.\ \ref{fig:potentials0}.
	Notice that for large values of $x$, the potentials $\epsilon_0(x)$, $\epsilon_1(x)$ and $V_k(x)$ become constant. In this region, we apply a CAP of the form
$
		W(x)	=	\alpha \left(x - x_{\text{CAP}} \right)^4  \cdot \Theta(x - x_{\text{CAP}}) , 
$
	with the Heaviside function $\Theta$, $\alpha=5$~eV/\AA$^4$ and $x_{\text{CAP}} = 3.5$~\AA, which is also depicted in Fig.\ \ref{fig:potentials0}.  The parameters entering the CAP are determined by converging the observables of interest.
	{\color{black} In the following, we describe the surfaces in the wide band limit.}
	{\color{black} We note in passing, that under certain conditions, there can be issues related to employing the wide band limit in cases where the coupling to the environment depends on the nuclear DOFs.\cite{Dou2016}}
	
	We use the model system, Eq.\ (\ref{eq:Hamiltonian_results}), to describe two different physical phenomena, namely the desorption of an adsorbate from a surface (see Fig.\ \ref{fig:setups}a), and the current-induced bond rupture in a single-molecule junction (see Fig.\ \ref{fig:setups}b). We recently used a similar model system to study current-induced dissociation in molecular junctions based on a mixed quantum classical methodology.\cite{Erpenbeck2018} For a detailed presentation and discussion of the model and the potentials, we refer to Ref.\ \onlinecite{Erpenbeck2018}. Qualitatively similar potentials were used to describe the interaction of molecules with surfaces.\cite{Miao2017}

	The results presented in the following are obtained by propagating Eq.\ (\ref{eq:EQM_nth_tier}) using a Runge-Kutta scheme. Thereby, initially, the electronic {\color{black} orbital} is unpopulated and the nuclear DOF is in the ground-state of $\epsilon_0(x)$. The temperature of the bath is $300$ K. All data are converged with respect to the number of DVR points, the number of tiers taken into account, and the number of Pade poles.

	First, we study a simple model for the desorption dynamics of a molecule from a surface as a function of time and coupling strength $\Gamma$.
	Here, $\Gamma$ is used as a scale for the molecule-surface coupling and is related to $\overline V_{k}$ from Eq.\ (\ref{eq:def_mol_lead_coupling_strength}) as 	$\overline V_{k} = \sqrt{\Gamma/2\pi}$.
	The corresponding setup is depicted in Fig.\ \ref{fig:setups}a; the desorption probability as a function of time for different values of $\Gamma$ is shown in Fig.\ \ref{fig:potentials}a. 
	Thereby, the desorption probability is defined as the population of the state ${x_\infty}$, which corresponds to the part of the wavefunction absorbed by the CAP.
	\begin{figure*}[tb!]
		\centering
		\begin{subfigure}[l]{0.325\textwidth}
			\raggedright \hspace*{1em} a)\\
			\includegraphics[width=0.9\textwidth]{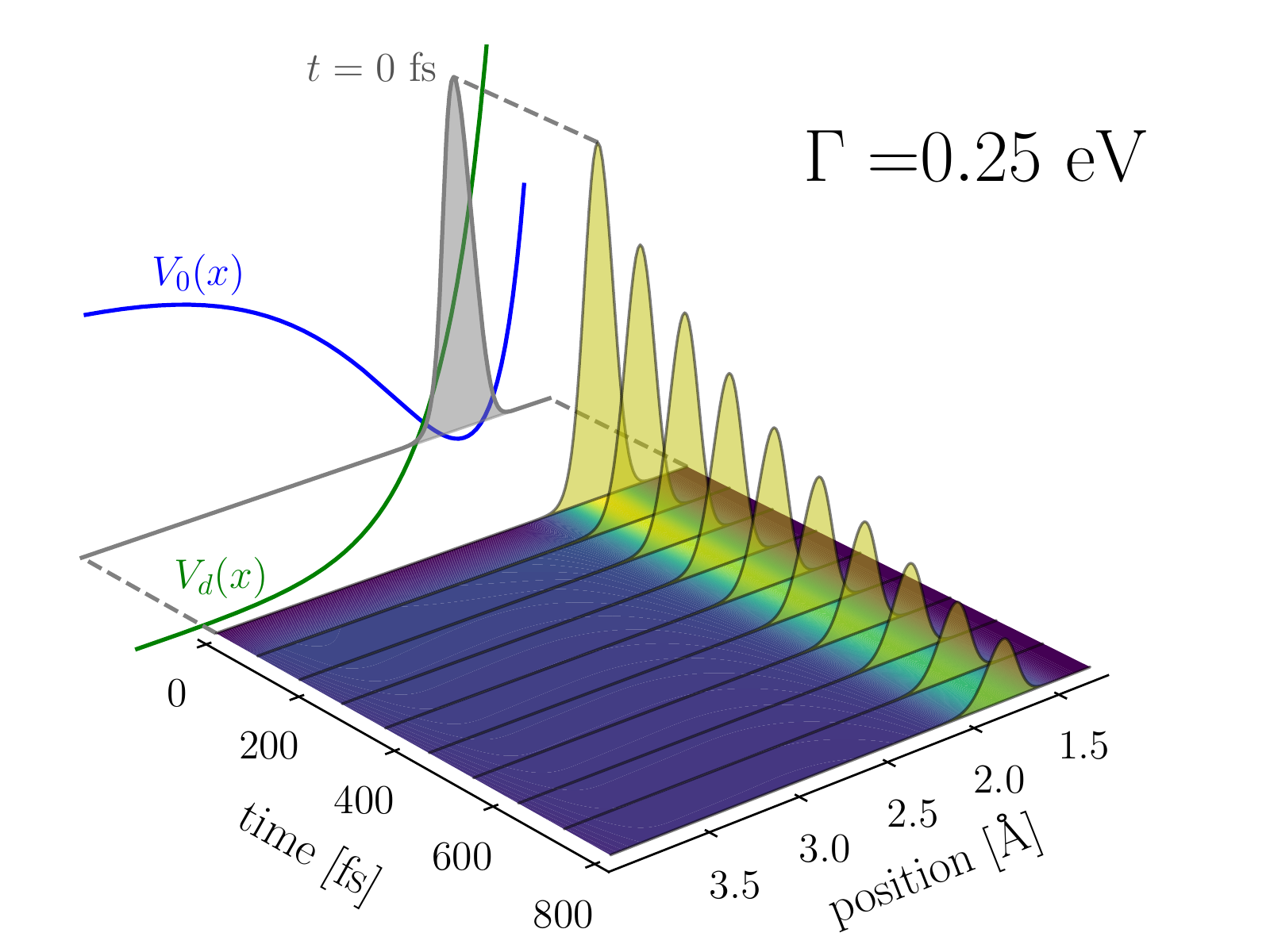}
		\end{subfigure}
		\begin{subfigure}[l]{0.325\textwidth}
			\raggedright \hspace*{1em} b) \\
			\includegraphics[width=0.9\textwidth]{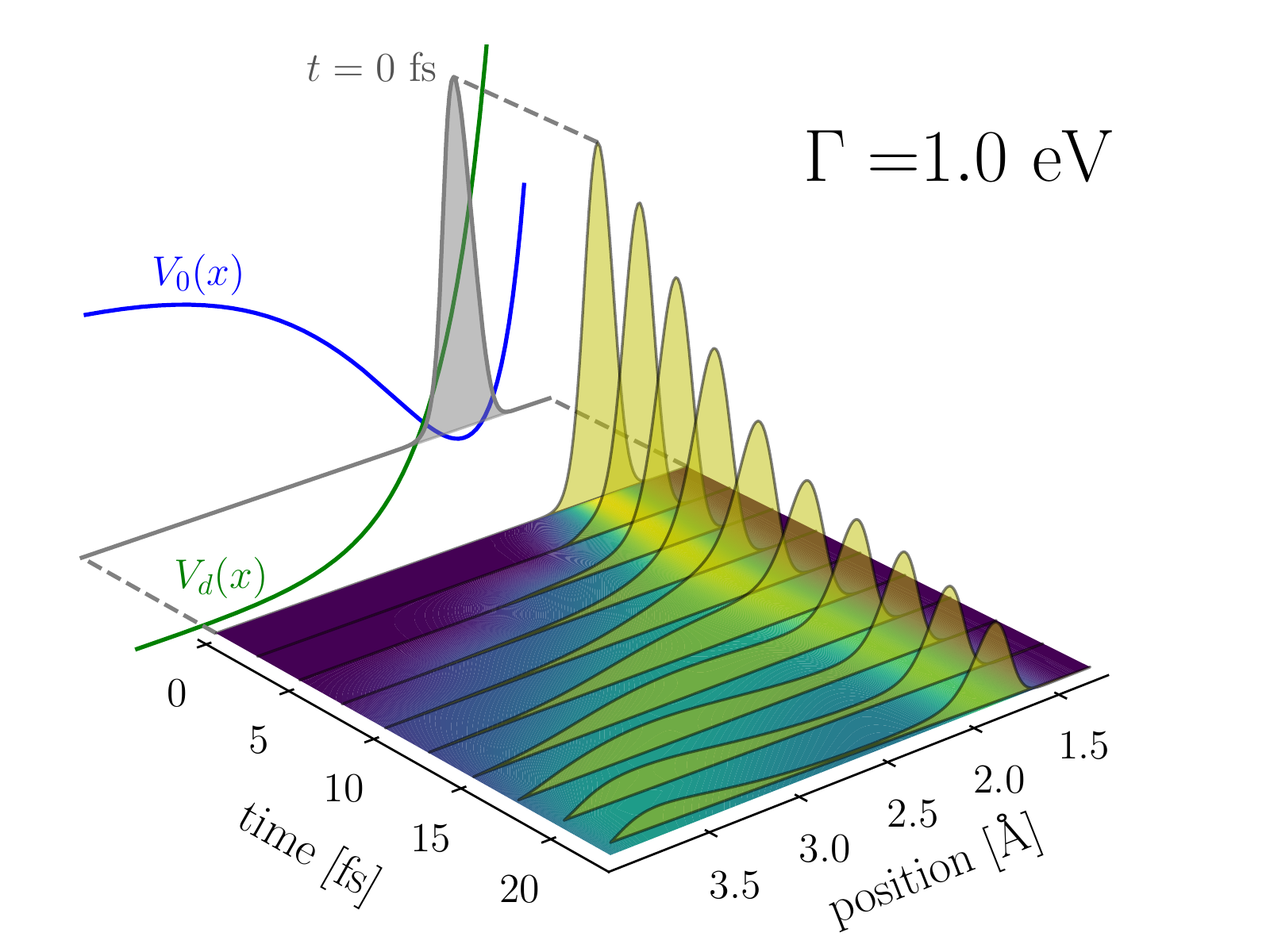}
		\end{subfigure}
		\begin{subfigure}[l]{0.325\textwidth}
			\raggedright \hspace*{1em} c)\\
			\includegraphics[width=0.9\textwidth]{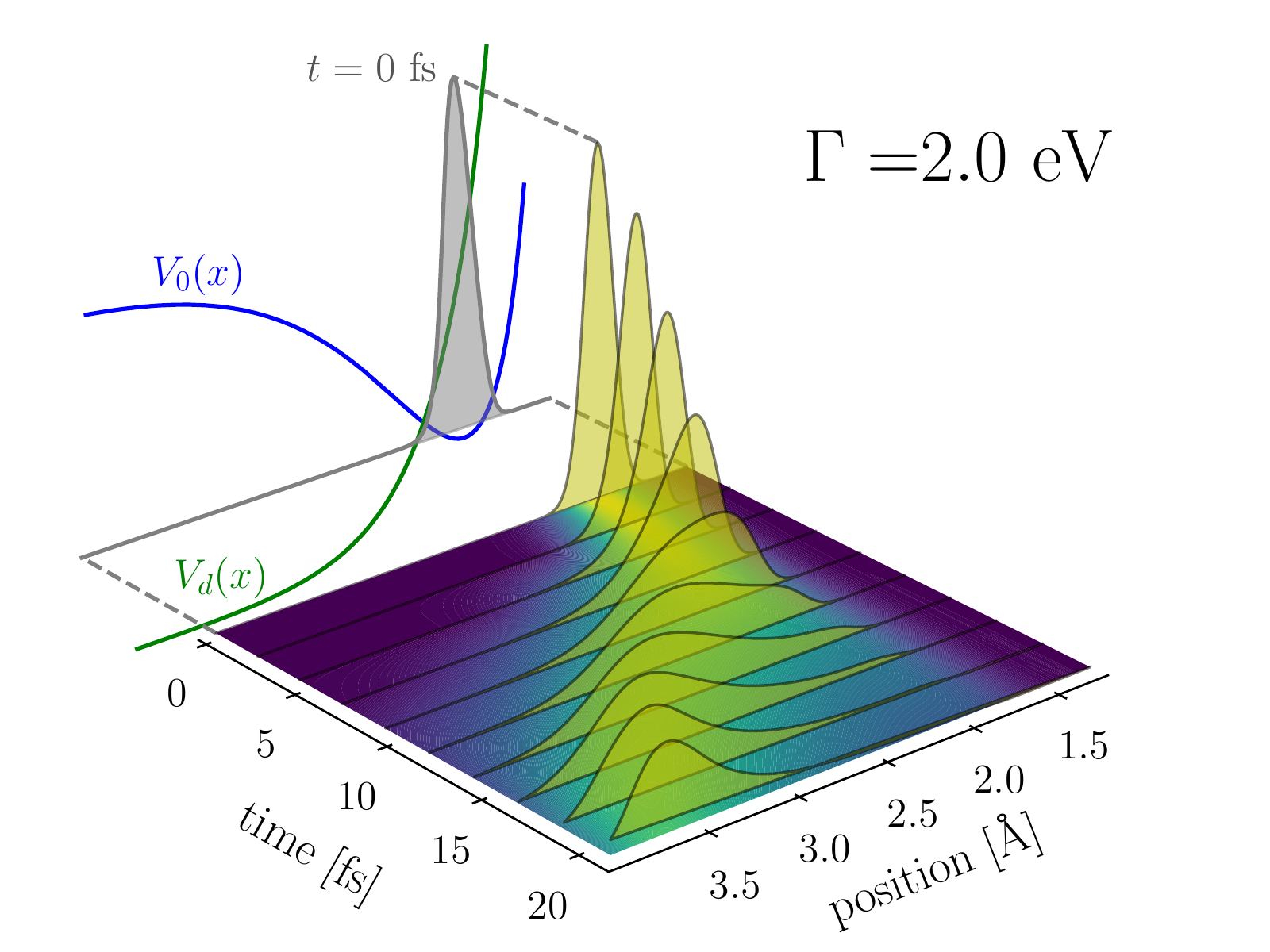}
		\end{subfigure}
		\captionsetup{font=small,labelfont=bf, justification=centerlast, format=plain}
		\caption{ \bf \scriptsize 
			Nuclear probability distribution as a function of time for different coupling strengths $\Gamma$.
			The yellow shaded areas represent the nuclear probability distribution for different times, the color grading at the bottom of the plot emphasizes the dynamics of the wave-packet.
			The initial state and the potentials are depicted in the background of the plots. 
			{\color{black} The barrier in the adiabatic potential of mean force is $\sim0.276$ eV ($\Gamma=0.25$ eV) and $\sim0.05$ eV ($\Gamma=1.0$ eV). There is no barrier for $\Gamma=2.0$ eV.}
		}
		\label{fig:wavepacket}
	\end{figure*}
	Fig.\ \ref{fig:potentials}a shows that the timescale for the desorption process changes over several orders of magnitude with the coupling strength $\Gamma$. This substantial change in the desorption dynamics can be explained by the different physical phenomena leading to desorption. 
	
	To this end, consider the dynamics of the reaction mode as a function of time for the three representative coupling strengths $\Gamma=0.25$ eV, $1$ eV and $2$ eV in Fig.\ \ref{fig:wavepacket}.
	For small $\Gamma$ as shown in Fig.\ \ref{fig:wavepacket}a, the overall shape of the wave-packet remains constant while the amplitude decreases. The corresponding desorption mechanism is based on tunneling of the nuclear reaction mode.
	For large $\Gamma$, desorption is mediated by the charging of the molecule. In this case, depicted in Fig.\ \ref{fig:wavepacket}c, the nuclear dynamics is quasi-classically. These limiting cases can be described by different approximate theories. In the small $\Gamma$ regime, low-order perturbative master equation approaches are applicable\cite{Egorova2003, Mitra2004, Leijnse2008}, whereas the adiabatic limit for large $\Gamma$ can be described by a classical Langevin equation approach.\cite{Lue2012, Dou2017, Miao2017, Dou2018}
	The HQME method can treat both limiting cases and is also applicable in the intermediate coupling regime where neither of the approximate methods is valid. In the latter regime, depicted in Fig.\ \ref{fig:wavepacket}b, the nuclear density splits into one part which remains located around the minimum of the neutral PES, and another part which propagates away from the surface describing the dissociative channel.
	
	As a second example to illustrate the novel methodology, we consider the current-induced dissociation of a chemical bond in a single-molecule junction in a setup depicted in Fig.\ \ref{fig:setups}b. Again, we employ the model defined in Eqs.\ (\ref{eq:Hamiltonian_results})--(\ref{eq:def_mol_lead_coupling_strength}) but now including two electronic baths, the left and right electrode.
	{\color{black} The applied bias voltage $\Phi$ is modeled as a symmetric shift in the chemical potentials, $\mu_{\text{L}}=-\mu_{\text{R}}=\Phi/2$. In the minimalistic model employed here, the molecular energy level is considered to be independent of the applied bias.}
	Fig.\ \ref{fig:potentials}b shows the dissociation dynamics for different bias voltages and weak molecule-surface coupling ($\Gamma=0.1$ eV). The results reveal a strong influence of the applied bias voltage. 
	For low bias voltages, the limited energy of the electrons entering the molecule results in a slow dissociation process. For large bias voltages, the higher energy of the electrons causes a large population of the molecular electronic {\color{black}orbital}. As a result, the nuclear wave-packet propagates in a quasi-classical manner on the anti-bonding PES resulting in a fast dissociation process.
	Because the molecule-electrode coupling in our model decreases upon dissociation (cf.\ Eq.\ (\ref{eq:def_mol_lead_coupling_strength})), the fast dissociation also results in a fast decrease of the current accross the junctions, especially for large voltages, which is seen in Fig.\ \ref{fig:potentials}c.\\

\paragraph*{\textbf{Conclusion:}}\label{sec:conclusion}

In this paper, we have introduced a novel method to simulate the quantum dynamics of a molecule coupled to one or several metal surfaces. The method combines the  HQME approach with a DVR representation for the nuclear DOFs to allow for the treatment of anharmonic PESs, which is essential for describing reaction dynamics at surfaces. Furthermore, a CAP in conjunction with {\color{black} an associated} source term is used to mimic an extended system with a finite DVR grid.
Being based on the HQME approach, the method is {\color{black}capable of providing} numerically exact results for a range of models, including the coupling to the continuum of electronic states of the surface and nonadiabatic processes. The effect of electronic friction, which is particularly important at metal surfaces, is included in a nonperturbative and non-Markovian way.

The numerical results presented in this paper highlight the applicability of the method to different problems such as the desorption of a molecule from a metal surface and current-induced rupture of bonds in molecular junctions.
It may provide benchmark results for the further advancement of mixed quantum classical and perturbative density-matrix schemes.

In the present implementation, the methodology is restricted to models with few reaction modes, which are treated in the system subspace. This limitation may be circumvented employing a reaction-surface Hamiltonian approach,\cite{Carrington1984} which describes additional, nonreactive modes within the harmonic approximation. Within the HQME method, these modes can then be treated efficiently within the bath subspace.
The method is also applicable to simulate photo-induced processes at metal surfaces by including the coupling to the light field into the system Hamiltonian.\\

\paragraph*{\textbf{Acknowledgement:}}

We thank Y.\ Ke, U.\ Peskin, and C.\ Schinabeck for helpful discussions. This work was supported by the German
Research Foundation (DFG).

\bibliography{Bib}

\end{document}